\documentclass{svjour3}                     % onecolumn (standard format)

\usepackage{graphicx}
\usepackage{natbib}

\usepackage{epsfig}

\newcommand\aap{A\&A}
\newcommand\apj{{ApJ}}

\journalname{SSRv}

\begin{document}

\title{Particle acceleration mechanisms}

\author{V. Petrosian \and
        A.M. Bykov}

\institute{V.Petrosian \at Department of Physics, Stanford
University, Stanford, CA 94305 \\
\email{vahep@stanford.edu} 
\and
A.M.Bykov \at A.F. Ioffe Institute for Physics and Technology,
          St. Petersburg, 194021, Russia\\
                  \email{byk@astro.ioffe.ru}}

\date{Received: 22 November 2007; Accepted: 14 december 2007}

\maketitle

\begin{abstract}
In this paper we review the possible mechanisms for production
of non-thermal electrons which are responsible for the observed
non-thermal radiation in clusters of galaxies. Our
primary focus is on non-thermal Bremsstrahlung
and inverse Compton scattering, that produce hard X-ray emission.
We first give a brief review of acceleration
mechanisms and point out that in most astrophysical situations,
and in particular for the intracluster medium, shocks, turbulence and
plasma waves play a crucial role. We also outline how the effects
of the turbulence can be accounted for. Using a generic model for
turbulence and acceleration, we then consider two scenarios for
production of non-thermal radiation. 
The first is motivated by the possibility that hard X-ray
emission is due to non-thermal Bremsstrahlung by nonrelativistic
particles and  attempts to produce non-thermal tails by
accelerating the electrons from the background plasma with an
initial Maxwellian distribution.  For acceleration rates smaller
than the Coulomb energy loss rate, the effect of energising the
plasma  is to primarily heat the plasma  with little sign of a
distinct non-thermal tail. Such tails are discernible only for
acceleration rates comparable or larger than the Coulomb loss
rate.  However, these tails are accompanied by significant
heating and they are present for a short  time of $<10^6$ yr, which
is also the time that the tail will
be thermalised. A longer period of acceleration at such rates will
result in a runaway situation with most particles being
accelerated to very high energies. These more exact treatments
confirm the difficulty with this model, first pointed out by
Petrosian (2001). Such non-thermal tails, even if possible, can only
explain the hard X-ray but not the radio emission which needs GeV
or higher energy electrons. For these and for production of hard
X-rays by the inverse Compton model, we need the second scenario
where there is injection and subsequent acceleration of
relativistic electrons. It is shown that a steady state situation,
for example arising from secondary electrons produced from cosmic
ray proton scattering by background protons, will most likely lead
to flatter than required electron spectra or it requires a short
escape time of the electrons from the cluster. An episodic
injection of relativistic electrons, presumably from galaxies or
AGN, and/or episodic generation of turbulence and shocks by
mergers can result in an electron spectrum consistent with
observations but for only a short period of less than one billion
years.
\end{abstract}

\section{Introduction}
\label{intro2}

In this paper we attempt to constrain the acceleration models based on the
observations described in the  papers by \citet{Durret2008},
\citet{Rephaeli2008} and \citet{Ferrari2008} - Chapters $4-6$, this volume, and
the required spectrum of the accelerated electrons shown in Fig.~6 of
\citealt{Petrosian2008} - Chapter 10, this volume. However, before addressing
these details  we first compare various acceleration processes and stress the
importance of plasma waves or turbulence (PWT) as an agent of scattering and
acceleration, and then describe the basic scenario and equations for treatment
of these processes. As pointed out  below there is growing evidence that PWT
plays an important role in acceleration of particles in general, and in clusters
of galaxies in particular. The two most commonly used acceleration mechanisms
are the following.

\subsection{Electric field acceleration}  

Electric fields parallel to magnetic fields can accelerate charged particles and
can arise as a result of magnetic field reconnection in a current sheet or other
situations. For fields less than  the so-called Dreicer field, defined as
$E_{\rm D} = {\rm k}T/({\rm e}\lambda_{\rm Coul})$, where
\begin{equation}
\lambda_{\rm Coul}\sim 15\ {\rm kpc}\left(\frac{T}{10^8\,{\rm
K}}\right)^{2}\left(\frac{10^{-3}\,{\rm cm}^{-3}}{n}\right)
\label{meanfp}
\end{equation}
is the collision (electron-electron or proton-proton) mean free
path\footnote{The proton-proton or ion-ion mean free path will be slightly
smaller because of the larger value of the Coulomb logarithm $\ln \Lambda\sim
40$ in the ICM.}, the rate of acceleration is less than the rate of collision
losses and only a small fraction of the particles can be accelerated into a
non-thermal tail of energy $E<L{\rm e}E_{\rm D}$. For the ICM $E_{\rm D}\sim
10^{-14}$~V\,cm$^{-1}$ and $L\sim 10^{24}$~cm so that sub-Dreicer fields  can
only accelerate particles up to 100's of keV, which is far below the  10's of
GeV electrons required by observations.  Super-Dreicer fields, which seem to be
present in many simulations of reconnection 
\citep{Drake2006,Cassak2006,Zenitani2005},  accelerate particles at
a rate that is faster than the collision or thermalisation time $\tau_{\rm
therm}$. This can lead to a runaway and an unstable electron distribution which,
as shown theoretically, by laboratory experiments and by the above mentioned
simulations, most probably will give rise to PWT  \citep{Boris1970,Holman1985}. 

{\sl In summary the electric fields arising as a result of reconnection cannot
be the sole agent of acceleration in the ICM, because there are no large scale
magnetically dominated cosmological flows, but it may locally produce an
unstable particle momentum distribution which will produce PWT that can then
accelerate particles.}

\subsection{Fermi acceleration} 

Nowadays this process has been divided into two kinds. In the original Fermi
process particles of velocity $v$ moving along magnetic field lines (strength
$B$) with a pitch angle $\cos\mu$ undergo random scattering by moving agents
with a velocity $u$. Because the head (energy gaining) collisions are more
probable than trailing (energy losing) collisions, on average, the particles
gain energy at a rate proportional to $(u/v)^2D_{\mu\mu}$, where $D_{\mu\mu}$ is
the pitch angle diffusion rate. This, known as a {\sl second order Fermi
process} is what we shall call stochastic acceleration. In general, the most
likely agent for scattering is PWT. An alternative process is what is commonly
referred to as a {\sl first order Fermi process}, where the actual acceleration
occurs when particles cross a  shock or any region of converging flow. Upon
crossing the shock  the fractional gain of momentum $\delta p/p\propto u_{\rm
sh}/v$. Ever since the 1970's, when several authors demonstrated that a very
simple version of this process leads to a power law spectrum that agrees
approximately with observations of the cosmic rays, shock acceleration is
commonly invoked in space and astrophysical plasmas. However, this simple model,
though very elegant, has some shortcomings specially when applied to electron
acceleration in non-thermal radiating sources. Moreover,  some of the features
that make this scenario for acceleration of cosmic rays attractive are not
present in most radiating sources where one needs efficient acceleration of
electrons to relativistic energies from a low energy reservoir.

The original, test particle theory of diffusive shock acceleration (DSA),
although very elegant and independent of geometry and other details
(e.g. \citealt{Blandford1978}) required several conditions such as injection of seed
particles and of course turbulence. A great deal of work has gone into
addressing these aspects of the problem and there has been a great deal of
progress. It is clear that nonlinear effects  (see e.g.
\citealt{Drury1983,Blandford1987,Jones1991,Malkov2001}) and losses (specially
for electrons)  play an important role and modify the resultant spectra and
efficiency of acceleration. Another important point is the source of the
turbulence or the scattering agents. A common practice is to assume Bohm
diffusion (see e.g. \citealt{Ellison2005}). Second order acceleration effects
could modify the particle spectra accelerated by shocks (see e.g.
\citealt{Schlickeiser1993,Bykov2000}). Although there are indications that
turbulence may be generated by the shocks and the accelerated particle upstream,
many details (e.g. the nature and spectrum of the turbulence) need to be
addressed more quantitatively. There has been progress on the understanding of
generation of the magnetic field and turbulence on strong shocks
\citep{Bell2001,Amato2006,Vladimirov2006} as required in recent observations of
supernova remnants (see e.g. \citealt{Volk2005}). There is also some evidence
for these processes from observations of heliospheric shocks (see e.g.
\citealt{Kennel1986,Ellison1990}). Basic features of particle acceleration by
cosmological shocks were discussed by \citealt{Bykov2008a} - Chapter 7, this
volume,  so we will concentrate here on the stochastic acceleration perspective.

\subsection{Stochastic acceleration} 

The PWT needed for scattering can also accelerate particles stochastically with
a rate $D_{EE}/E^2$, where $D_{EE}$ is the energy diffusion coefficient, so that
shocks may not be always necessary.  In low beta plasmas, $\beta_{\rm
p}=2(v_{\rm s}/v_{\rm A})^2<1$, where the  Alf\'ven velocity $v_{\rm
A}=\sqrt{B^2/4\pi \rho}$, the sound velocity $v_{\rm s}=\sqrt{{\rm k}T/m}$, 
$\rho=nm$ is the mass density and $n$ is the number density of the gas, and for
relativistic particles the PWT-particle interactions are dominated by Alf\'venic
turbulence, in which case the rate of energy gain $D_{EE}/E^2=(v_{\rm
A}/v)^2D_{\mu\mu}\ll D_{\mu\mu}$, so that the first order Fermi process is more
efficient. However, at low energies and/or in very strongly magnetised plasmas,
where $v_{\rm A}$  can exceed ${\rm c}$, the speed of light\footnote{Note that
the Alf\'ven group velocity $v_{\rm g}= {\rm c}\sqrt{v_{\rm A}^2/(v_{\rm
A}^2+{\rm c}^2)}$ is always less than ${\rm c}$.}, the acceleration rate may
exceed the scattering rate (see \citealt{Pryadko1997}), in which case low energy
electrons are accelerated more efficiently by PWT than by shocks.\footnote{In
practice, i.e. mathematically, there is little difference between the two
mechanisms \protect\citep{Jones1994}, and the acceleration by turbulence and 
shocks can be combined (see below).}

Irrespective of which process dominates the particle acceleration, it is clear
that PWT has a role in all of them. Thus, understanding of the production of PWT
and its interaction with particles  is extremely important. Moreover, turbulence
is expected to be present in most astrophysical plasmas including the ICM and in
and around merger or accretion shocks, because the ordinary and magnetic
Reynolds numbers  are  large. Indeed turbulence may be the most efficient
channel of energy dissipation. In recent years there has been a substantial
progress in the understanding of MHD turbulence
\citep{Goldreich1995,Goldreich1997,Lithwick2003,Cho2002,Cho2006}. These provide
new tools for a more quantitative investigation of turbulence and the role it
plays in many astrophysical sources.

\section{Turbulence and stochastic acceleration}
\label{turb}

\subsection{Basic scenario}
\label{scenario}

The complete picture of stochastic acceleration by PWT is a complex and not yet
fully understood or developed process. However, one  might  envision the
following scenario.

Turbulence or plasma waves can be generated in the ICM on some macroscopic scale
$L \sim$ 300 kpc (some fraction of the cluster size or some multiple of galactic
sizes) as a result of merger events or by accretion or merger shocks. That these
kind of motions or flows with velocity comparable to or somewhat greater than
the virial velocity $u_L\sim$ 1000 km\,s$^{-1}$ will lead to PWT is very likely,
because in the ICM  the ordinary Reynolds number $R_{\rm e}=u_LL/\nu\gg1$. Here
$\nu\sim v_{\rm th}\lambda_{\rm scat}/3$ is the viscosity, $v_{\rm
th}=\sqrt{{\rm k}T/m}\sim u_L(T/10^8)^{1/2}$ and $\lambda_{\rm scat}$ is the
mean free path length. The main uncertainty here is in the value of
$\lambda_{\rm scat}$. For Coulomb collisions $\lambda_{\rm scat}\sim 15$ kpc
(Eq.~\ref{meanfp}) and $R_{\rm e}\sim 100$ is just barely large enough for
generation of turbulence. However, in a recent paper \citet{Brunetti2007} argue
that in the presence of a magnetic field of $B\sim\mu$G, $v_{\rm A}\sim 70
(B/\mu{\rm G})(10^{-3}\,{\rm cm}^{-3}/n)^{1/2}$ km\,s$^{-1}$ is much smaller
than $v_{\rm th}$ so that the turbulence will be super-Alfv\'enic, in which case
the mean free path may be two orders of magnitude smaller\footnote{Plasma
instabilities, possibly induced by the relativistic particles, can be another
agent of decreasing the effective particle mean free path
\protect\citep{Schekochihin2005}.} yielding $R_{\rm e}\sim 10^4$. We know this
also to be true from a phenomenological consideration. In a cluster the hot gas
is confined by the gravitational field of the total  (dark and 'visible')
matter. Relativistic particles, on the other hand, can cross the cluster of
radius $R$ on a timescale of $T_{\rm cross} = 3\times 10^6\,(R/$Mpc)~yr and can
escape the cluster  (see Fig.~\ref{timescales} below), unless confined by a
chaotic magnetic field or a scattering agent such as turbulence with a mean free
path $\lambda_{\rm scat} \ll R$. If so, then the escape time $T_{\rm esc}\sim
T_{\rm cross}(R/\lambda_{\rm scat})=T_{\rm cross}^2/\tau_{\rm scat}$. The curve
marked with arrows in this figure shows the maximum value of the required
$\tau_{\rm scat}$ so that the escape time is longer than the energy loss time
$\tau_{\rm loss}$. As is evident from this figure, for a GeV electron to be 
confined for a Hubble timescale, or $T_{\rm esc}\sim 10^{10}$~yr, we need
$\tau_{\rm scat}\sim 3\times 10^4$~yr or $\lambda_{\rm scat} < 10$~kpc. This
could be the case in a chaotic magnetic field and/or in the presence of
turbulence. Some observations related to this are discussed by
\citealt{Petrosian2008} - Chapter 10, this volume; see also \citet{Vogt2005}.
Numerous numerical simulations also agree with this general picture. There is
evidence for large scale bulk flows in the simulations of merging clusters (e.g.
\citealt{Roettiger1996,Ricker2001}), and that these are converted into
turbulence with energies that are a substantial fraction of the thermal energy
of the clusters (e.g. \citealt{Sunyaev2003,Dolag2005}).  For more details see
\citet{Brunetti2007}.

Once the PWT is generated it can undergo two kind of interactions. The first is
dissipationless cascade from wave vectors $k_{\min}\sim L^{-1}$ to smaller
scales. The cascade is gouverned by the rates of {\sl wave-wave interactions.}
For example, in the case of weak turbulence, that can be considered as a
superposition of weakly interacting wave packets, the three wave interactions
can be represented as 
\begin{equation} 
\label{wave-wave}
\omega({\bf k}_1) + \omega({\bf k}_2) = \omega({\bf k}_3) \,\,\,\,
{\rm and} \,\,\,\, {\bf k}_1+{\bf k}_2={\bf k}_3, 
\end{equation}
where ${\bf k}$ is the wave vector, and the wave frequency, $\omega({\bf k})$,
is obtained from the {\sl plasma dispersion relation}.  One can interpret
Eq.~\ref{wave-wave} as energy-momentum conservation laws for weakly coupled
plasma waves in a close analogy to the optical waves. The interaction rates can
be represented by the wave diffusion coefficient $D_{ij}$ or the cascade time
$\tau_{\rm cas}\sim k^2/D_{ij}$. The largest uncertainty is in the diffusion
coefficient. Because of the nonlinear nature of the interactions this
coefficient depends on the wave spectrum $W({\bf k})$. As mentioned above there
has been considerable progress in this area in the past two decades and there
are some recipes how to calculate the diffusion coefficients.

The second  is {\bf damping of the PWT} by {\sl wave-particle
interaction} which terminates the dissipationless cascade, say at
an outer scale $k_{\max}$ when the damping rate
$\Gamma(k_{\max})=\tau_{\rm cas}^{-1}(k_{\max})$. The range
$k_{\min}<k<k_{\max}$ is called the inertial range. The damping rate
can be obtained from the finite temperature dispersion relations
(see below). The energy lost from PWT goes into heating the
background plasma and/or accelerating particles into a non-thermal
tail. These processes are described by  the diffusion coefficients
$D_{EE}$ and $D_{\mu\mu}$ introduced above. These coefficients are
obtained from consideration of the wave-particle interactions
which are often dominated by resonant interactions, specially for
low beta (magnetically dominated) plasma, such that
\begin{equation}
\label{resonance} 
\omega({\bf k})-k\cos\theta v\mu = n\Omega/\gamma, 
\end{equation}
for waves propagating at an angle $\theta$ with respect to the large scale
magnetic field, and a particle of velocity $v$, Lorentz factor $\gamma$, pitch
angle $\cos\mu$ and gyrofrequency $\Omega={\rm e}B/m{\rm c}$. Both cyclotron
(the term in the right hand side of Eq.~\ref{resonance}) and Cerenkov resonance
(the second term in the left hand side) play important roles in the analysis
(see for details e.g. \citealt{Akhiezer1975}). Here, when the harmonic number
$n$ (not to be confused with the density) is equal to zero, the process is
referred to as the {\sl transit time damping}. For gyroresonance damping by
waves propagating parallel to the field lines ($\theta=0$) $n=\pm 1$. For
obliquely propagating waves, in principle one gets contributions from all
harmonics $n= \pm 1,\  \pm 2,\ \ldots$, but for practical purposes most of the
contribution comes from the lowest harmonics $n= \pm 1$ (see
\citealt{Pryadko1998}).

\subsection{Dispersion relations}

It is clear from the above description that at the core of the evaluation of
wave-wave or wave-particle interactions (and all the coefficients of the kinetic
equations described below) lies the plasma dispersion relation $\omega({\bf
k})$. It describes the characteristics of the waves that can be excited in the
plasma, and the rates of wave-wave and wave-particle interactions.

\begin{figure}
\includegraphics[height=0.85\textwidth,angle=270]{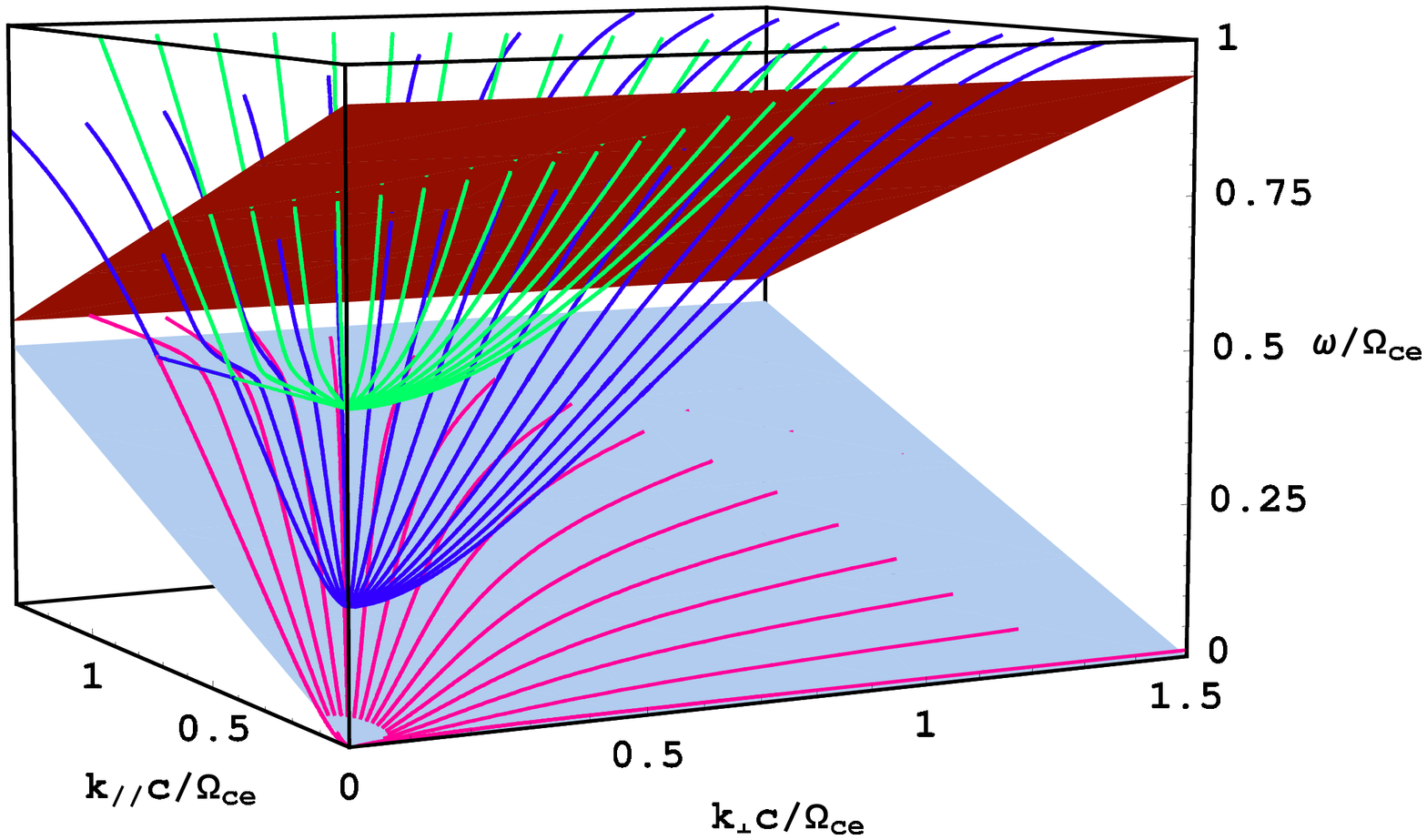}\\
\includegraphics[height=0.85\textwidth,angle=270]{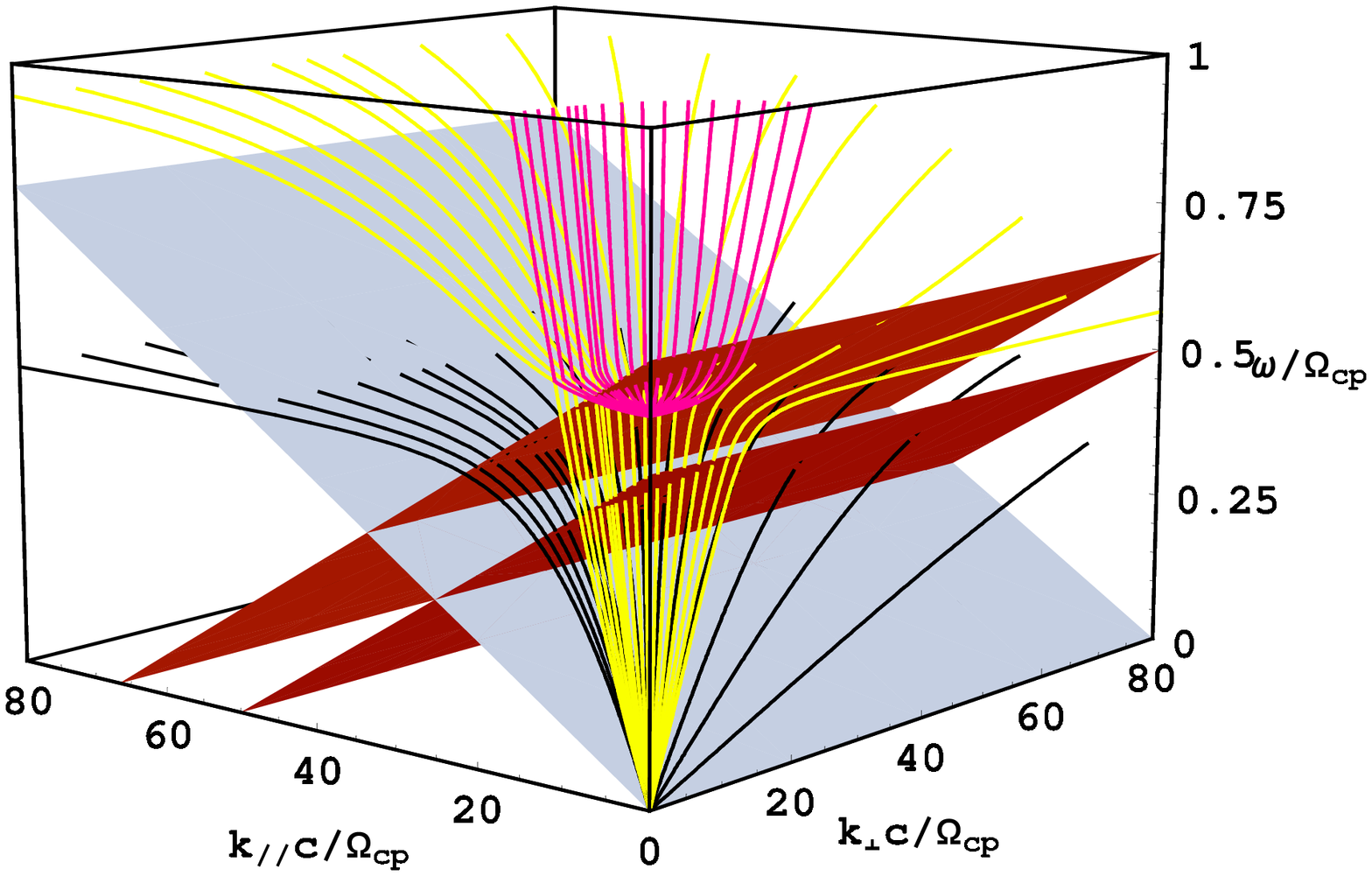}
\caption{Dispersion relation (curves) surfaces for
a {\sl cold} fully ionised H and He (10~\% by number) plasma
and resonance condition (flat) surfaces showing the regions around
the electron (top panel) and proton (bottom panel) gyro-frequencies. Only
waves with positive $k_\|, k_\perp$ (or $0<\theta<\pi/2$) are shown.
The mirror image with respect to the ($\omega,\ k_{\perp}$) plane
gives the waves propagating in the opposite direction. From high
to low frequencies, we have one of the electromagnetic branches
(green), upper-hybrid branch (purple), lower-hybrid branch, which
also includes the whistler waves (pink), fast-wave branches
(yellow), and Alfv\'{e}n branch (black). The effects of a finite
temperature modify these curves at frequencies $\omega\sim kv_{\rm
th}$, where $v_{\rm th}=\sqrt{2{\rm k}T/m}$ is the thermal velocity (see
e.g. \protect\citealt{Andre1985}). 
The resonance surfaces are for electrons with
$v=0.3{\rm c}$ and $|\mu|=1.0$ (top panel: upper, brown $n=1$, lower, light
blue  $n=0$) and $^4$He (bottom panel: middle, brown $n=1$) and $^3$He
(bottom panel: upper, brown $n=1$) ions with $|\mu|=1.0$ and $v=0.01{\rm c}$.
The resonance surfaces for the latter two are the same when $n=0$
(bottom panel: lower). } 
\label{dispersion}
\end{figure}

In the MHD regime for a cold plasma
\begin{equation}
\label{mhddisp}
\omega=v_{\rm A}k\cos\theta\,\,\,\,\,  {\rm and}\,\,\,\,\,  \omega=v_{\rm A}k
\end{equation}
for the Alfv\'en and the fast (magneto-sonic) waves, respectively. Beyond the
MHD regime a multiplicity of wave modes can be present and the dispersion
relation is more complex and is obtained from the following expressions (see
e.g. \citealt{Sturrock1994}):
\begin{equation}
\label{dispgeneral}
\tan^2\theta = {-P(n_r^2-R)(n_r^2-L)\over (Sn_r^2-RL)(n_r^2-P)},
\end{equation}
where $n_r=k{\rm c}/\omega$ is the refractive index, $S={1\over 2}(R+L)$, and
\begin{equation}
\label{dispterms}
P= 1-\sum_i{\omega_{pi}^2\over \omega^2},\ \ \  
R=1-\sum_i{\omega_{{\rm p}i}^2\over
\omega^2}\left({\omega\over \omega+\epsilon_i\Omega_i}\right),\ \ \ 
{\rm and} \ \ \
L= 1-\sum_i{\omega_{{\rm p}i}^2\over \omega^2}\left({\omega\over
\omega-\epsilon_i\Omega_i}\right).
\end{equation}
Here $\omega_{{\rm p}i}^2= 4\pi n_i q_i^2/ m_i$ and
$\Omega_i=|q_i|B / m_i {\rm c}$ are the plasma and gyro frequencies,
$\epsilon_i=q_i / |q_i|$, and $n_i$, $q_i$, 
and $m_i$ are the density, charge, and mass of the background particles. For
fully ionised plasmas such as that in the ICM it is sufficient to include terms
due to electron, proton and $\alpha$ particles. Fig.~\ref{dispersion} shows the
dispersion surfaces (depicted by the curves) obtained from the above expressions
along with the resonant planes in the ($\omega,\ k_\parallel,\ k_\perp$) space.
Intersections between the dispersion surfaces and the resonant planes define the
resonant wave-particle interactions and the particle kinetic equation
coefficients. One can also envision a similar graphic description of the three
wave interactions (Eq.~\ref{wave-wave}) using the intersections of the curved
dispersion surfaces. However, such calculations have been carried out only in
the MHD regime using the simple relations of Eq.~\ref{mhddisp}, which is already
a complicated procedure (see e.g. \citealt{Chandran2005,Luo2006}).

The above dispersion relations are good approximations for low beta plasmas but
in the ICM the plasma beta is large:
\begin{equation}
\label{beta}
\beta_{\rm p}=8\pi n{\rm k}T/B^2=3.4\times 10^{2}(n/10^{-3}\,
{\rm cm}^{-3})(\mu{\rm G}/B)^2(T/10^8\,{\rm K}).
\end{equation}
For high beta plasmas the dispersion relation is modified, specially for higher
frequencies $\omega\sim kv_{\rm th}$. For example, in the MHD regime, in
addition to the Alfv\'en mode one gets fast and slow modes with the dispersion
relation (see e.g. \citealt{Sturrock1994})
\begin{equation}
\label{mhddisp2}
(\omega/k)^2={1\over 2}\left[(v_{\rm A}^2+v_{\rm s}^2)\pm \sqrt{v_{\rm
A}^4+v_{\rm s}^4-2v_{\rm A}^2v_s^2\cos {2\theta}}\right],
\end{equation}
and the more general dispersion relation (Eq.~\ref{dispgeneral})  is modified in
a more complicated way (see e.g. \citealt{Andre1985} or \citealt{Swanson1989}). 
The finite temperature imparts an imaginary part $\omega_{\rm i}$ to the wave
frequency that gives the damping rate $\Gamma(k)$ as long as  $\omega_{\rm
i}<\omega_{\rm r}$, the real part of the frequency\footnote{Note that the
'thermal' effects change $\omega_{\rm r}$ only slightly so that often the real
part, the resonant interaction rate and the particle diffusion coefficients can
be evaluated using the simpler cold plasma dispersion relation depicted in 
Fig.~\ref{dispersion}.}.  For more details see e.g. 
\citet{Barnes1973,Swanson1989,Pryadko1998,Pryadko1999,Cranmer2003,Brunetti2007}. 
In general, these rates and the modification of the dispersion relation are
known for Maxwellian (sometimes anisotropic) energy distributions of the plasma
particles. For non-thermal distributions the damping rates can be evaluated as
described \citet{Petrosian2006} using the coupling described in
Eq.~\ref{coefficients} below.

\subsection{Kinetic equations and their coefficients}
\label{kinetic}

\subsubsection{Wave equation} 

Adopting the diffusion approximation (see e.g. \citealt{Zhou1990}), one can
obtain the evolution of the spatially integrated wave spectrum $W({\bf k}, t)$
from the general equation
\begin{equation}
\\{\partial {W} \over \partial t}
= {\partial \over\partial k_i}\left[D_{ij}{\partial\over \partial
k_j}{W}\right] - \Gamma({\mathbf k}){W} - {{W}\over T^{W}_{\rm
esc}({\mathbf k})} + \dot{Q}^{W}, 
\label{waves} 
\end{equation} 
where $\dot{Q}^{\rm W}$ is the rate of generation of PWT at $k_{\min}$, $T^{\rm
W}_{\rm esc}$ is the escape time, and $D_{ij}$ and $\Gamma$ describe the cascade
and damping of the waves. The calculation of the damping rate is complicated but
as described above it is well understood, but there are many uncertainties about
the treatment of the  cascade process or the form of $D_{ij}$. This is primarily
because of incompleteness of the theoretical models and sufficient observational
or experimental data. There are some direct observations in the Solar wind (e.g.
\citealt{Leamon1998}) and indirect inferences in the interstellar medium ( see
e.g. \citealt{Armstrong1995}). There is some hope \citep{Inogamov2003} of future
observations in the ICM. Attempts in fitting the Solar wind data have provided
some clues about the cascade diffusion coefficients (see
\citealt{Leamon1999,Jiang2007}).

\subsubsection{Particle acceleration and transport} 

As described by \citealt{Petrosian2008} - Chapter 10, this volume, the general
equation for treatment of particles is the Fokker-Planck equation which for ICM
conditions can be simplified considerably. As pointed out above we expect a
short mean free path and fast scatterings for all particles. When the scattering
time $\tau_{\rm scat}=\lambda_{\rm scat}/v\sim \langle 1/D_{\mu\mu}\rangle$ is
much less than the dynamic and other timescales, the particles will have an
isotropic pitch angle distribution. The pitch-angle averaged and spatially
integrated particle distribution is obtained from\footnote{The derivation of
this equation for the stated conditions and some other details can be found in
the Appendix.}  
\begin{equation} 
{\partial N(E,t) \over
\partial t}
 =  {\partial \over \partial E}\left[D_{EE}{\partial N(E,t) \over \partial E}
 - (A-\dot E_L) N(E,t)\right]
 - {N(E,t) \over T^{\rm p}_{\rm esc}} +{\dot Q}^{\rm p}.
\label{KEQ}
\end{equation}
Here $D_{EE}/E^2$ is  the energy diffusion, due to scattering by PWT as
described above and due to Coulomb collisions as discussed by
\citealt{Petrosian2008} - Chapter 10, this volume, $A(E)/E\sim \zeta
D_{EE}/E^2$, with $\zeta(E)=(2-\gamma^{-2})/(1+\gamma^{-1})$  is the rate of 
direct acceleration due to interactions with PWT and all other agents, e.g.,
direct first order Fermi acceleration by shocks, ${\dot E}_L/E$ is the energy
loss rate of the particles (due to Coulomb collisions and synchrotron and IC
losses, see Fig.~\ref{timescales} in \citealt{Petrosian2008} - Chapter 10, this
volume),  and $\dot{Q}^{\rm p}$ and the term with the escape times $T^{\rm
p}_{\rm esc}$ describe the source and leakage of particles\footnote{in what
follows we will assume that the  waves are confined to  the ICM so that 
$T^{W}_{\rm esc}\rightarrow\infty$ and in some cases we will assume no escape of
particles and let $T^{\rm p}_{\rm esc}\rightarrow\infty$.}.

The above two kinetic equations are coupled by the fact that the coefficients of
one depend on the spectral distribution of the other; the damping rate of the
waves depends on $N(E,t)$ and the diffusion and accelerations rates of particles
depend on the wave spectrum $W({\bf k}, t)$. Conservation of energy requires
that the energy lost by the waves ${\dot {\cal W}_{\rm tot}}\equiv \int
\Gamma({\bf k})W({\bf k}){\rm d}^3k$ must be equal to the energy gained by the
particles from the waves;  ${\dot {\cal E}}=\int [A(E)-A_{\rm sh}]N(E){\rm d}
E$. Representing the energy transfer rate between the waves and particles by
$\Sigma ({\bf k}, E)$ this equality implies that
\begin{equation}
\label{coefficients}
\Gamma({\bf k})= \int\limits_0^\infty {\rm d}E\, N(E)\Sigma({\bf k}, E),\,\,\,\,
A(E)=\int\limits_0^\infty {\rm d}^3kW({\bf k})\Sigma ({\bf k}, E)
+ A_{\rm sh},
\end{equation}
where we have added $A_{\rm sh}$ to represent contributions of other
(non-stochastic acceleration) processes affecting the direct acceleration, e.g.,
shocks.

If the damping due to non-thermal particles is important then the wave and
particle kinetic equations (\ref{waves}) and (\ref{KEQ}) are coupled and
attempts have been made to obtain solutions of the coupled equations
\citep{Miller1996,Brunetti2005}. However, most often the damping rate is
dominated by the background thermal particles so that the wave and non-thermal
particle kinetic equations decouple. This is a good approximation in the ICM
when dealing with relativistic electrons so that for determination of the
particle spectra all we need is the boundaries of the inertial range ($k_{\min},
k_{\max}$), the wave spectral index $q$ in this range (most likely $5/3<q<3/2)$,
and the shape of the spectrum above $k_{\max}$ which is somewhat uncertain (see
\citealt{Jiang2007}).

\section{Particle acceleration in clusters of galaxies}

We now address the problem of particle acceleration in clusters of galaxies. The
current information on the ICM does not allow us to treat the problem as
outlined above by solving the coupled kinetic equations. In what follows we make
reasonable assumptions about the turbulence and the particle diffusion
coefficients, and then solve the particle kinetic equation to determine
$N(E,t)$. We first consider the apparently simple scenario of acceleration of
the background thermal particles. Based on some general arguments, \citet[P01
hereafter]{Petrosian2001}  showed that this is not a viable mechanism. Here we
carry out a more accurate calculation and show that this indeed is the case.
This leads us to consider the transport and acceleration of high energy
particles injected into the ICM by other processes.

\subsection{Acceleration of background particles}

\begin{figure}
\leavevmode\centering
\includegraphics[width=120mm,height=120mm]{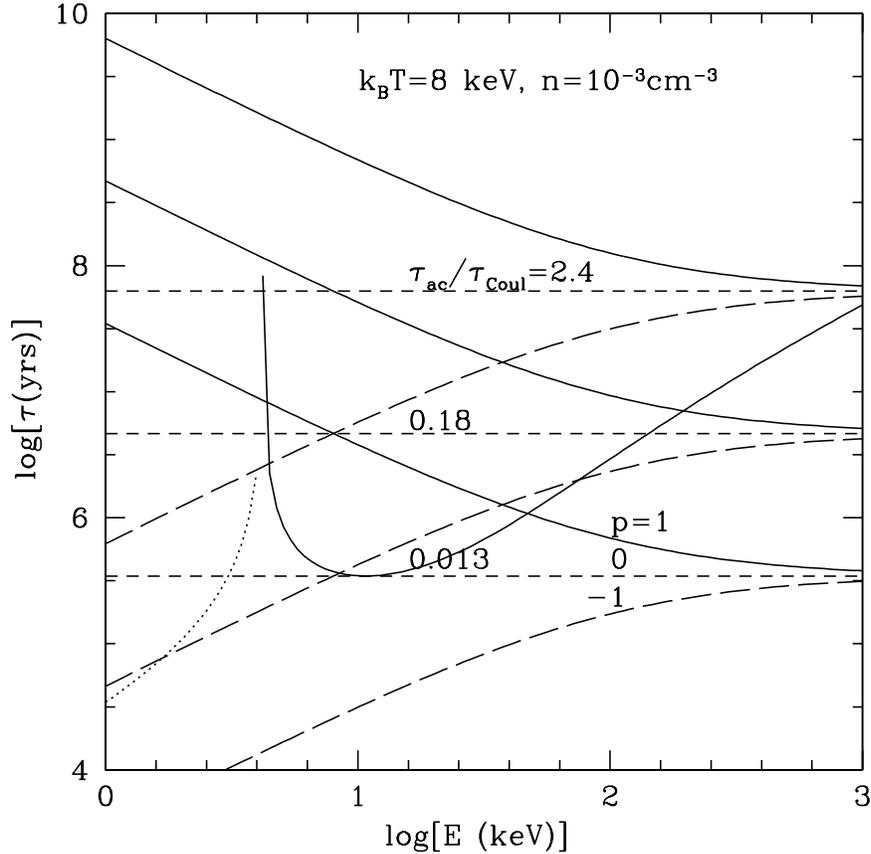}
\caption{\footnotesize Acceleration and loss timescales for ICM conditions based
on the model described in the text. We use the effective Coulomb loss rate given
by \protect\citealt{Petrosian2008} - Chapter 10, this volume, 
and the IC plus synchrotron losses for a CMB temperature
of $T_{\rm CMB}=3$ K and an ICM magnetic field of $B=1$~$\mu$G. We also use
the simple acceleration scenario of Eq.~\ref{acctime} for
 $E_{\rm c}=0.2m_{\rm e}{\rm c}^2$ ($\sim 100$~keV) and for the three
specified values of $p$ and times $\tau_0/\tau_{\rm Coul}$ (from 
\citealt{Petrosian2007}).}
\label{timescales1}
\end{figure}

The source particles to be accelerated are the ICM hot electrons subject to
diffusion in energy space by turbulence and Coulomb collisions, acceleration by
turbulence or shocks, and energy losses due to Coulomb collisions\footnote{In
our numerical results we do include synchrotron, IC and Bremsstrahlung losses.
But these have an insignificant effect in the case of nonrelativistic electrons
under investigation here.}. We start with an ICM of ${\rm k}T=8$~keV,
$n=10^{-3}$~cm$^{-3}$ and assume a continuous injection of turbulence so that
its density remains constant resulting in a time independent diffusion and
acceleration rate. The results described below is from a recent paper by
\citet[PE07 hereafter]{Petrosian2007}. Following this paper we assume a simple
but generic energy dependence of these coefficients.  Specifically we assume a
simple acceleration rate or timescale
\begin{equation} 
\label{acctime}
\tau_{\rm ac}=E/A(E)=\zeta D(E)/E^2=\tau_0(1+E_{\rm c}/E)^p.
\end{equation}
Fig.~\ref{timescales1} shows a few examples of these time scales along with
the effective Coulomb (plus IC and synchrotron) loss times as described in
Fig.~3 of \citealt{Petrosian2008} - Chapter 10, this volume.

\begin{figure}
\begin{center}
\includegraphics[width=0.49\textwidth]{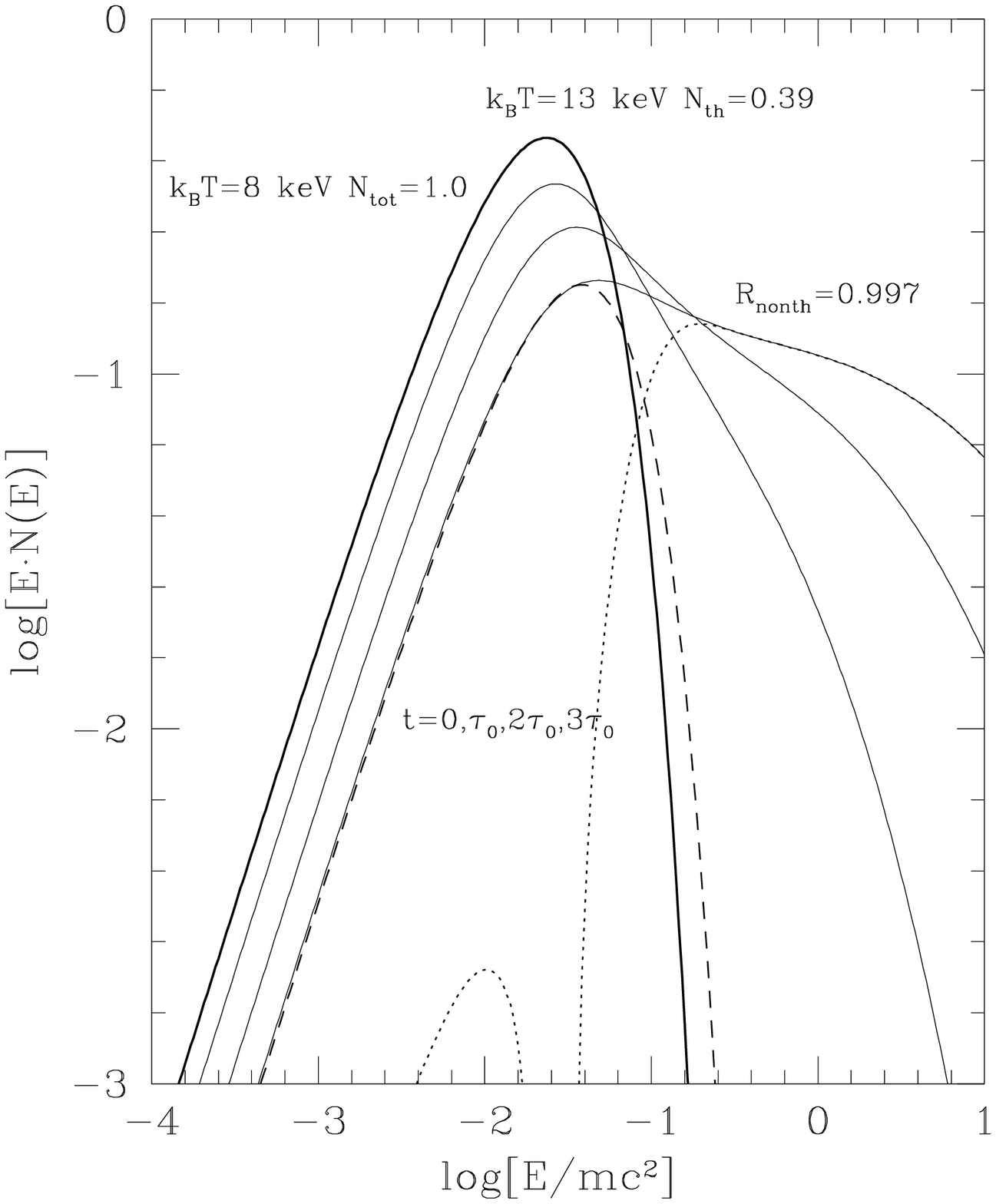}
\includegraphics[width=0.49\textwidth]{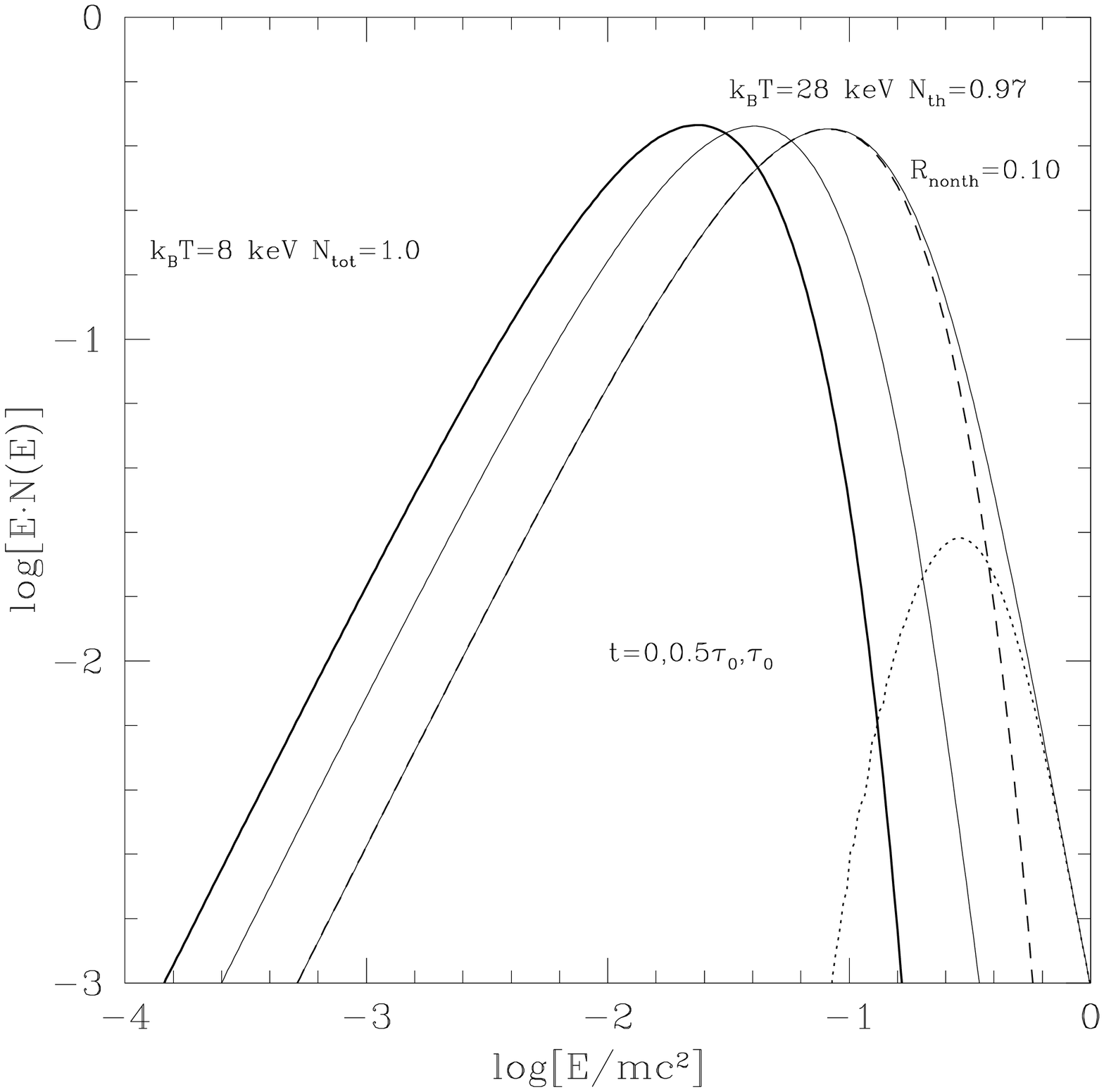}\\
\includegraphics[width=0.49\textwidth]{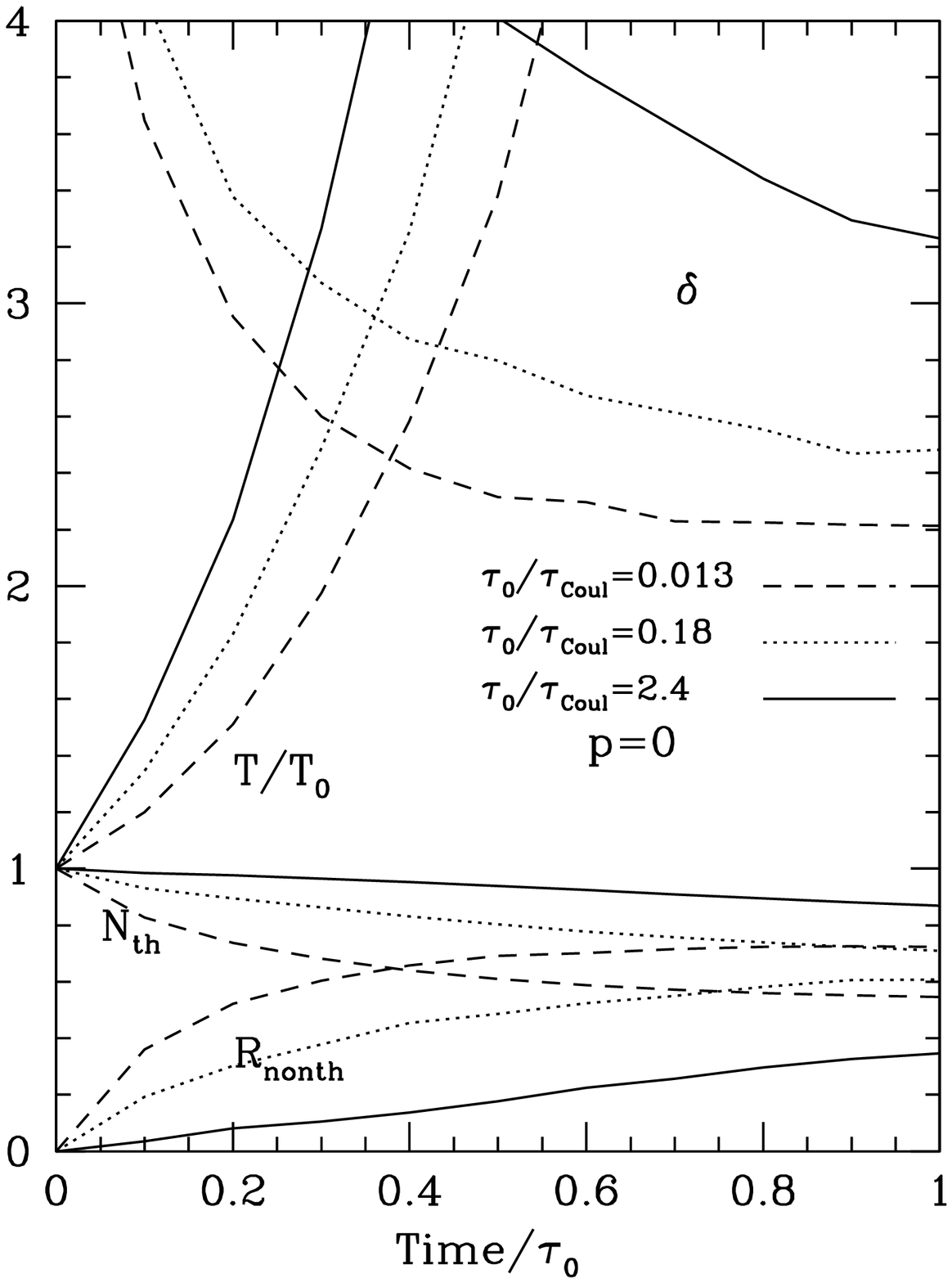}
\end{center}
\caption{{\sl Upper left panel:} Evolution with time of electron spectra in the
presence of a constant level of turbulence that accelerates electrons according
to  Eq.~\ref{acctime} with $\tau_0/\tau_{\rm Coul}=0.013$,  $E_{\rm c}=0.2$
($\sim 100$~keV) and $p=1$. For the last spectrum obtained for time $t=\tau_0$,
the low end of the spectrum is fitted to a thermal component (dashed curve). The
residual 'non-thermal' part is shown by the dotted curve.  We also give the
initial and final values of the temperature, the fraction of electrons in the
thermal component $N_{\rm th}$, and the ratio of energy of the non-thermal
component to the thermal components $R_{\rm nonth}$. 
{\sl Upper right panel:} Same as above except for $\tau_0/\tau_{\rm Coul}=2.4$. Note
that now there is only heating and not much of acceleration.
{\sl Lower panel:} Evolution with time (in units of $\tau_0$) of electron
spectral parameters, $T(t)/T_0, N_{\rm th}, R_{\rm
nonth}$ and the power-law index $\delta$ for indicated values of
$\tau_0/\tau_{\rm Coul}$  and for $p=0$ and $E_{\rm c}=100$ keV. Note that
for models with the same value of $p$ at $t=\tau_0$ roughly the same amount of
energy has been input into the ICM (from \citealt{Petrosian2007}).}
\label{spectra}
\end{figure}

We then use Eq.~\ref{KEQ} to obtain the time evolution of the particle spectra.
After each time step we use the resultant spectrum to update the Coulomb
coefficients as described by \citealt{Petrosian2008} - Chapter 10, this volume. 
At each step the electron spectrum can be divided into a quasi-thermal and a
'non-thermal' component.  A best fit Maxwellian distribution to the 
quasi-thermal part is obtained, and we determine a temperature and the fraction
of the thermal electrons. The remainder is labelled as the non-thermal tail.
(For more details see PE07). The left and middle panels of Fig.~\ref{spectra}
show two spectral evolutions for two  different values of acceleration time
$\tau_0/\tau_{\rm Coul} = 0.013$ and 2.4, respectively, and for  $E_{\rm
c}=25$~keV and $p=1$. The last spectrum in each case is for time $t=\tau_0$,
corresponding to an equal energy input for all cases.  The initial and final
temperatures, the fraction of particles in the quasi-thermal component $N_{\rm
th}$, and the ratio of non-thermal to thermal energies $R_{\rm nonth}$ are shown
for each panel. The general feature of these results is that the turbulence
causes both acceleration and heating in the sense that the spectra at low
energies resemble a thermal distribution but also have a substantial deviation
from this quasi-thermal distribution at high energies which can be fitted by a
power law over a finite energy range.  The distribution is broad and continuous,
and  as time progresses it becomes broader and shifts to higher energies; the
temperature increases and the non-thermal 'tail' becomes more prominent. There
is very little of a non-thermal tail for $\tau_0>\tau_{\rm Coul}$ and most of
the turbulent energy goes into heating (middle panel). Note that this also means
that for a steady state case where the rate of energy gained from turbulence is
equal to radiative energy loss rate (in this case thermal Bremsstrahlung, with
time scale $\gg\tau_{\rm Coul}$) there will be an insignificant non-thermal
component.  There is no distinct non-thermal tail except at unreasonably high
acceleration rate (left panel). Even here there is significant heating (almost
doubling of the temperature) within a short time ($\sim 3\times 10^5$~yr). At
such rates of acceleration most particles will end up at energies much larger
than the initial ${\rm k}T$ and in a broad non-thermal distribution.  We have
also calculated spectra for different values of the cutoff energy $E_{\rm c}$
and index $p$. As expected for larger (smaller) values of $E_{\rm c}$ and
smaller (higher) values of $p$ the fraction of non-thermal particles is lower
(higher).

The evolution in time of the temperature (in units of its initial value), the
fraction of the electrons in the 'non-thermal' component, the energy ratio
$R_{\rm nonth}$ as well as an index $\delta=-{\rm d}\,\ln N(E)/{\rm d}\,\ln E$
for the non-thermal component are shown in the right panel of
Fig.~\ref{spectra}. All the characteristics described above are more clearly
evident in this panel and similar ones for $p=-1$ and +1. In all cases the
temperature increases by more than a factor of 2. This factor is smaller at
higher rates of acceleration. In addition, high acceleration rates produce
flatter non-thermal tails (smaller $\delta$) and a larger fraction of
non-thermal particles (smaller $N_{\rm th}$) and energy ($R_{\rm nonth}$).

It should be noted that the general aspects of the above behaviour are dictated
by the Coulomb collisions and are fairly insensitive to the details of the
acceleration mechanism which can affect the spectral evolution somewhat
quantitatively but not its qualitative aspects. At low acceleration rates one
gets mainly heating and at high acceleration rate a prominent non-thermal tail
is present but there is also substantial heating within one acceleration
timescale which for such cases is very short. Clearly in a steady state
situation there will be an insignificant non-thermal component. These findings
support qualitatively findings by P01 and do not support the presence of
distinct non-thermal tails advocated by \citet{Blasi2000} and \citet{Dogiel2007},
but agree qualitatively with the more rigorous analysis of \citet{Wolfe2006}.
For further results, discussions and comparison with earlier works see PE07.

{\sl We therefore conclude that the acceleration of background electrons
stochastically or otherwise and non-thermal bremsstrahlung are not a viable
mechanism for production of non-thermal hard X-ray excesses observed in some
clusters of galaxies.}

\subsection{Acceleration of injected particles}

\begin{figure}
\leavevmode\centering
\includegraphics[width=120mm,height=120mm]{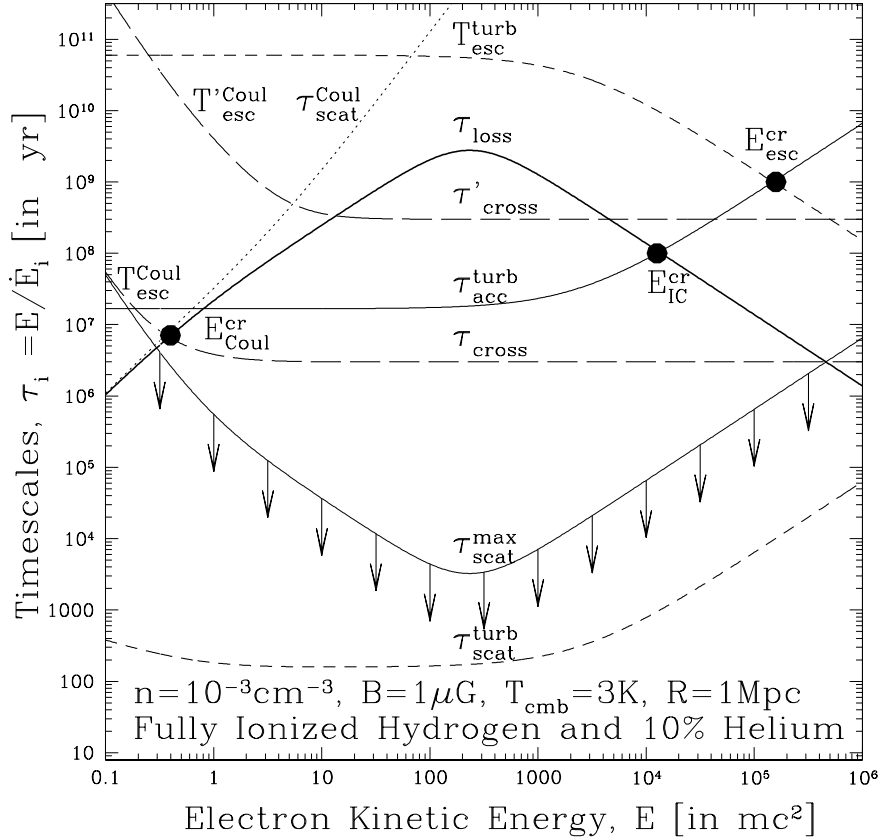}
\caption{Comparison of the energy dependence of the total loss time (radiative
plus Coulomb; thick solid line) with timescales for scattering $\tau_{\rm scat}$
(lower dashed line), acceleration $\tau_{\rm ac}$ (thin solid lines), crossing
time $\tau_{\rm cross} \sim R/c\beta$ (long dashed lines) and escape time
$T_{\rm esc}\sim \tau_{\rm cross}^2/\tau_{\rm scat}$ (upper dashed line) of
electrons. At low energies the scattering and escape times are dominated by
Coulomb scattering but at high energies turbulence scattering becomes more
important. The acceleration is constant at low energies ($A(E)\propto E$) but
increases at high energies, corresponding to an acceleration rate $A(E)
\rightarrow$ a constant at high $E$ or to a turbulence spectral index $q = 1$. A
chaotic magnetic field with scale of 10~kpc will increase the crossing time to
the primed curve. The arrows show the maximum scattering times for which the
escape times are equal or longer than the total loss times. The critical
energies where the  acceleration time is equal to the escape time,  the Coulomb
and inverse Compton loss times are shown.}
\label{timescales}
\end{figure}

The natural way to overcome the above difficulties is to assume that the radio
and the hard X-ray radiation are produced by relativistic electrons injected in
the ICM, the first via synchrotron and the second via the inverse Compton
scattering of CMB photons. The energy loss rate of relativistic electrons can be
approximated by (see P01)
\begin{equation}
\label{loss}
{\dot E}_L (E)/E_p=(1+(E/E_{\rm p})^2)/\tau_{\rm loss},
\end{equation}
where
\begin{equation}
\label{values}
\tau_{\rm loss}=E_{\rm p}/(4\pi {\rm r}_0^2m_{\rm e}{\rm c}^3n\ln\Lambda)\,\,\,\,\,
{\rm and}\,\,\,\,\,
E_{\rm p}\simeq m_{\rm e}{\rm c}^2\,[{9\over /8} 
{n\ln\Lambda\over u_{\rm ph}+B^2/8\pi}]^{1/2}
\end{equation}
are twice the loss time and the energy where the  total loss curve reaches its
maximum\footnote{We ignore the Bremsstrahlung loss and the weak dependence on
$E$ of Coulomb losses at nonrelativistic energies. We can also ignore the energy
diffusion rate due to Coulomb scattering.} (see Fig.~\ref{timescales}). Here
${\rm r}_0 = {\rm e}^2/(m_{\rm e}{\rm c}^2 ) = 2.82\times 10^{-13}$~cm is the
classical electron radius, $u_{\rm ph}$ (due to the CMB) and $B^2/8\pi$ are
photon (primarily CMB) and magnetic field energy densities. For the ICM
$B\sim\mu$G, $n = 10^{-3}$~cm$^{-3}$ and the Coulomb logarithm $\ln\Lambda=40$
so that $\tau_{\rm loss}=6.3\times 10^9$ yr and $E_{\rm p}=235 m_{\rm e}{\rm
c}^2$.

The electrons are scattered and gain energy if there is some turbulence in the
ICM. The turbulence should be such that it resonates with the injected
relativistic electrons and not the background thermal nonrelativistic electrons
for the reasons described in the previous section. Relativistic electrons will
interact mainly with low wavevector waves in the inertial range where
$W(k)\propto k^{-q}$ with the index $q\sim 5/3$ or 3/2 for a Kolmogorov or
Kraichnan cascade. There will be little interaction with nonrelativistic
background electrons if the turbulence spectrum is cut off above some maximum
wave vector $k_{\max}$ whose value depends on viscosity and magnetic field.  The
coefficients of the transport equation (Eq.~\ref{KEQ}) can then be approximated
by
\begin{equation}
\label{coef} 
D(E)={\cal D}E^{q}, A(E)=a{\cal D}E^{q-1},\,\,\,
{\rm and}\,\,\,T_{\rm esc}={\cal T}_{\rm esc} E^s. 
\end{equation}
For a stochastic acceleration model at relativistic energies $a=2$, but if in
addition to scattering by PWT there are other agents of acceleration (e.g.
shocks) then the coefficient $a$ will be larger than 2. In this model the escape
time is determined by the crossing time $T_{\rm cross}\sim R/c$ and the
scattering time $\tau_{\rm scat}\sim D_{\mu\mu}^{-1}$. We can then write $T_{\rm
esc}\sim T_{\rm cross}(1+T_{\rm cross}/\tau_{\rm scat})$. Some examples of these
are shown in Fig.~\ref{timescales}. However, the escape time is also affected by
the geometry of the magnetic field (e.g. the degree of its entanglement). For
this reason we have kept the form of the escape time to be more general. In
addition to these relations we also need the spectrum and rate of injection to
obtain the spectrum of radiating electrons. Clearly there are several
possibilities. We divided it into two categories: {\sl steady state} and {\sl
time dependent}. In each case we first consider only the effects of losses,
which means ${\cal D}=0$ in the above expressions, and then the effects of both
acceleration and losses.

\subsubsection{Steady state cases}

By steady state we mean variation timescales of order or larger than the Hubble
time which is also longer than the maximum loss time $\tau_{\rm loss}/2$. Given
a particle injection rate ${\dot Q}={\dot Q}_0 f(E)$ (with $\int f(E){\rm
d}E=1$) steady state is possible if ${\cal T}_{\rm esc}={\dot Q}_0/\int
N(E)E^{-s}{\rm d}E$.

In the absence of acceleration (${\cal D}=0$) eQ.~\ref{KEQ} can be solved
analytically. For the examples  of escape times given in  Fig.~\ref{timescales}
($T_{\rm esc}>\tau_{\rm loss}$) one gets the simple cooling spectra $N=({\dot
Q}\tau_{\rm loss}/E_{\rm p})\int_E^\infty f(E){\rm d}E/(1+(E/E_{\rm p})^2)$,
which gives a spectral index break at $E_{\rm p}$ from index $p_0-1$ below to
$p_0+1$ above $E_{\rm p}$, for an injected power law $f(E)\propto E^{-p_0}$. For
$p_0=2$ this will give a high energy power law in rough agreement with the
observations but with two caveats. The first is that the spectrum of the
injected particles must be cutoff below $E\sim 100 m_{\rm e}{\rm c}^2$ to avoid
excessive heating  and the second is that this scenario cannot produce the
broken power law or exponential cutoff we need to explain the radio spectrum of
Coma (see Fig.~6  and the discussion in \citealt{Petrosian2008} - Chapter 10,
this volume).  A break is possible only if the escape time is shorter than
$\tau_0$ in  which case the solution of the kinetic equation for a power law
injected spectrum ($p_0>1$ and $s>-1$) leads to the broken power law
\begin{equation}
\label{inside} 
N(E) = Q_0 \cases{{\cal T}_{\rm esc}(E/E_p)^{-p_0+s}
&if  $E\ll E_{\rm cr}$,\cr \tau_{\rm loss}(E/E_p)^{-p_0-1}/(s+1) &if
$E_{\rm cr}\ll E$, \cr} 
\end{equation}
where $E_{\rm cr}=E_{\rm p}((s+1)({\cal T}_{\rm esc}/\tau_{\rm
loss})^{-1/(s+1)}$. Thus, for $p_0\sim 3$ and $s=0$ and $T_{\rm esc}\simeq
0.02\tau_{\rm loss} $ we  obtain a spectrum with a break at $E_{\rm cr}\sim
10^4$, in agreement with the radio data (\citealt{Rephaeli1979} model). However,
this also means that  a large fraction of the $E<E_{\rm p}$ electrons escape
from the ICM, or more accurately from the turbulent confining region, with a
flux of $F_{\rm esc}(E) \propto N(E)/T_{\rm esc}(E)$. Such a short escape time
means a scattering time which is only ten times shorter than the crossing time
and a mean free path of about $\sim 0.1R\sim 100$ kpc. This is in disagreement
with the Faraday rotation observations which imply a tangled magnetic field
equivalent to a ten times smaller mean free path. The case for a long escape
time was first put forth by \citet{Jaffe1977}.

\begin{figure}
\leavevmode\centering
\includegraphics[width=0.48\textwidth]{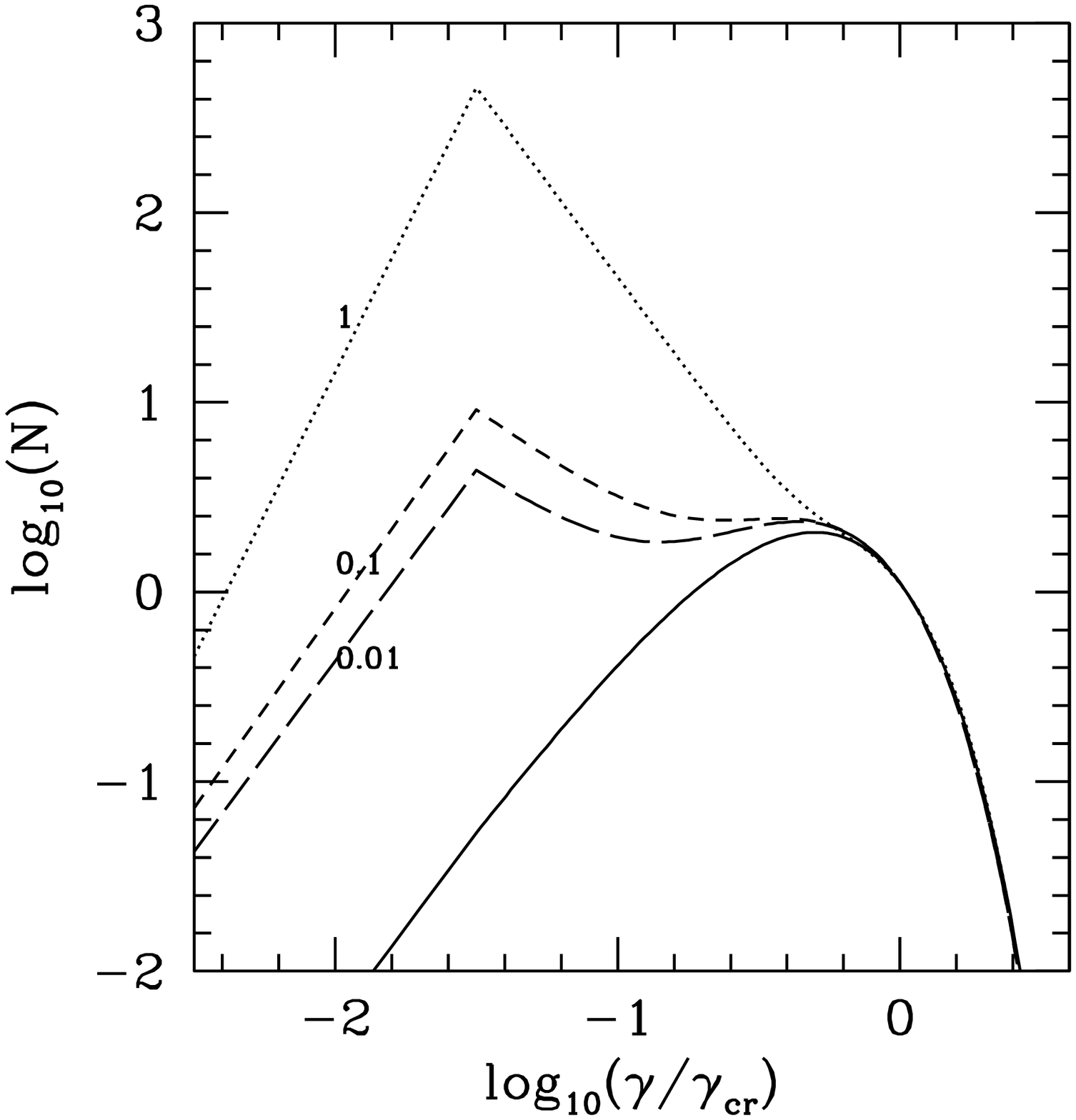}
\includegraphics[width=0.48\textwidth]{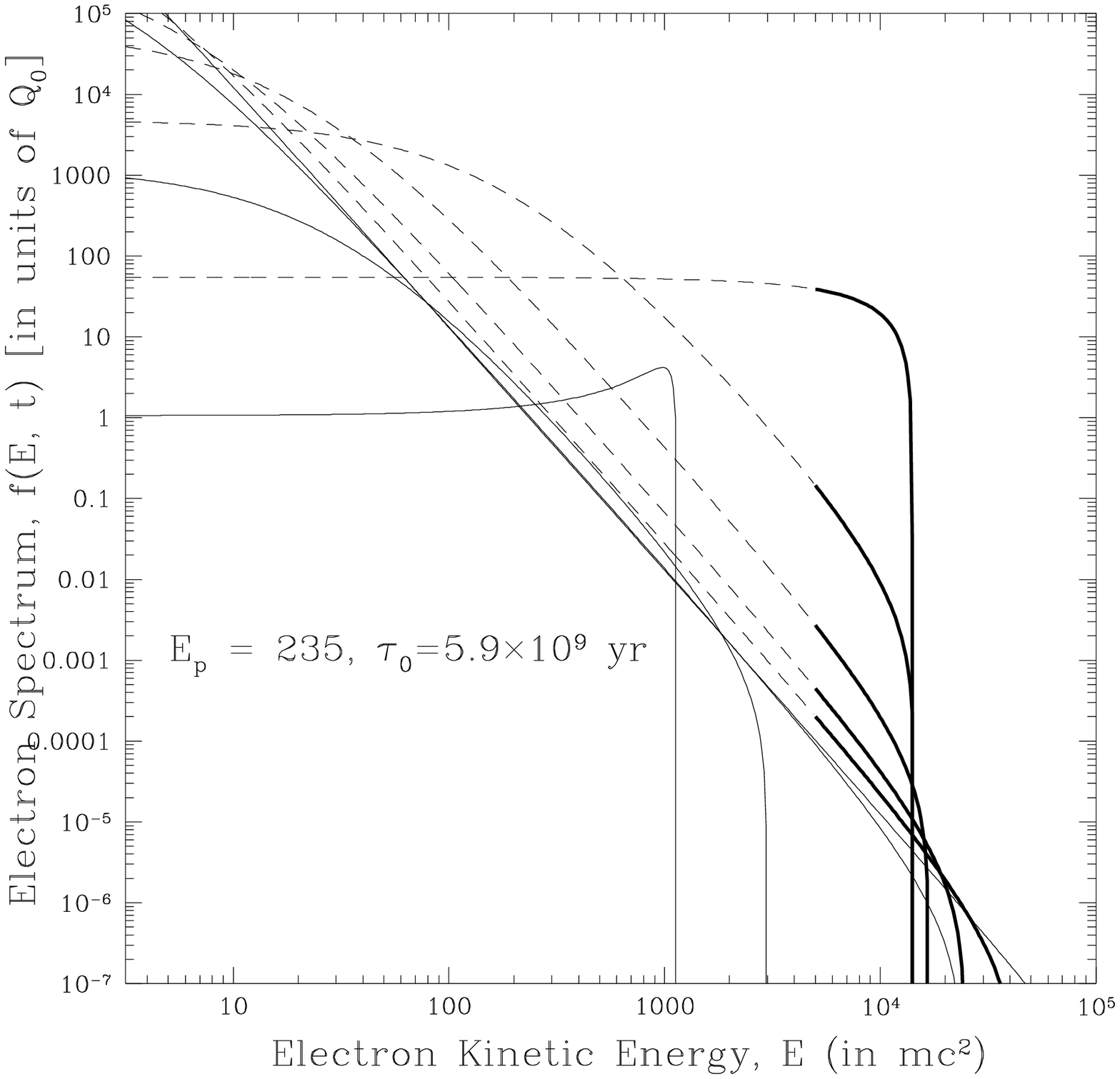}
\caption{{\sl Left panel:}  The steady state electron spectra injected at energy
$E_0$ and subject to continuous acceleration by turbulence with spectral index
$q=2$ for different values of the ratio $T_{\rm esc}/\tau_{\rm loss}$. Note that
$\tau_0$ in the label is the same as $\tau_{\rm loss}$ in the text. For very
high values of this ratio we get a relativistic Maxwellian distribution (from
Liu et al. 2006). {\sl Right panel:} Evolution with time of a power law injected
spectrum (top line) subject to Coulomb and inverse Compton (plus synchrotron)
losses as given by Eq.~\ref{prompt2}. Solid lines for $b=2$ ($t/\tau_0=
10^{-n}$; $n= 3$, 2, 1, 0), and dashed lines for $b=60$ ($t/\tau_0= 10^{-2.18 +
n/3}$; $n=0$, 1, 2, 3, 4). The heavy portions show the energy range of the
electrons needed for production of radio and hard X-rays.}
\label{spectra1}
\end{figure}

Thus it appears that in addition to injection of relativistic electrons we also
need a steady presence or injection of PWT to further scatter and accelerate the
electrons. The final spectrum of electrons will depend on the acceleration rate
and its energy dependence. In general, when the acceleration is dominant one
expects a power law spectrum. Spectral breaks appear at critical energies when
this rate becomes equal to and smaller than other rates such as the loss or
escape rates (see Fig.~\ref{timescales}). In the energy range where the losses
can be ignored electrons injected at energy $E_0$ ($f(E)=\delta(E-E_0)$) one
expects a power law above (and below, which we are not interested in) this
energy. In the realistic case of long $T_{\rm esc}$ (and/or when the direct
acceleration rate is larger than the rate of stochastic acceleration (i.e. $a\gg
1$) then spectral index of the electrons will be equal to $-q+1$ requiring a
turbulence spectral index of $q=4$ which is much larger than expected values of
5/3 or 3/2 (see \citealt{Park1995}). This spectrum will become steeper (usually
cut off exponentially) above the energy where the loss time becomes equal to the
acceleration time $\tau_{\rm ac} =E/A(E)$ or at $E_{\rm cr}=(E_{\rm p}a{\cal
D}\tau_{\rm loss})^{1/(3-q)}$. Steeper spectra below this energy are possible
only for shorter $T_{\rm esc}$. The left panel of Fig.~\ref{spectra1} shows the
dependence of the spectra on $T_{\rm esc}$ for $q=2$ and $s=0$ (acceleration and
escape times independent of $E$). The spectral index just above $E_0$ is
$p=\sqrt {9/4+2\tau_{\rm ac}/T_{\rm esc}}-1.5$. In the limit when $T_{\rm
esc}\rightarrow \infty$ the distribution approaches a relativistic Maxwellian
distribution $N\propto E^2{\rm e}^{-E/E_{\rm cr}}$. For a cut-off energy $E_{\rm
cr}\sim 10^4$ this requires an acceleration time of $\sim 10^8$ yr and for a
spectral index of $p=3$ below this energy we need $T_{\rm esc}\sim \tau_{\rm
ac}/18\sim 5\times 10^6$ yr which is comparable to the unhindered crossing time.
This is too short. As shown in Fig.~\ref{timescales} any scattering mean free
path (or magnetic field variation scale) less than the cluster size will
automatically give a longer escape time and a flatter than required spectrum.
For further detail on all aspects of this case see \citet{Park1995}, P01 and
\citet{Liu2006}.

{\sl In summary there are several major difficulties with the steady state
model.}

\subsubsection{Time dependent models}

We are therefore led to consider time dependent scenarios with time variation
shorter than the Hubble time.  The time dependence may arise from the episodic
nature of the injection process (e.g. varying AGN activity) and/or from episodic
nature of turbulence generation process (see e.g. \citealt{Cassano2005}).  In
this case we need solutions of the time dependent equation (Eq.~\ref{KEQ}).  We
start with the generic model of a prompt single-epoch injection of electrons with
$Q(E,t)=Q(E)\delta(t-t_0)$.  More complex temporal behaviour can be obtained by
the convolution of the injection time profile with the solutions described
below. The results presented below are from P01. Similar treatments of the
following cases can be found in \citet{Brunetti2001} and \citet{Brunetti2007}.

It is clear that if there is no re-acceleration, electrons will lose energy
first at highest and lowest energies due to inverse Compton and Coulomb losses,
respectively. Particles will be peeled away from an initial power law with the
low and high energy cut-offs moving gradually toward the peak energy $E_{\rm
p}$. A more varied and complex set of spectra can be obtained if we add the
effects of diffusion and acceleration.  Simple analytic solutions for the time
dependent case are possible only for special cases.  Most of the complexity
arises because of the diffusion term which plays a vital role in shaping the
spectrum for a narrow injection spectrum.  For some examples see
\citet{Park1996}. Here we limit our discussion to a broad initial electron
spectrum in which case the effects of this term can be ignored until such
features are developed. Thus, if we set $D(E)=0$, which is a particularly good
approximation when $a\gg 1$, and for the purpose of demonstration if we again
consider the simple case of constant acceleration time ($q=2$ and $A(E)=a{\cal
D}E$), then the solution of Eq.~\ref{KEQ} gives
\begin{equation}
\label{prompt2}
N(E,t) = \exp \{-t/T_{\rm esc}\}Q_0{{[T_+ - (E/E_{\rm p})\tan(\delta
t/\tau_{\rm loss})/\delta]^{p_0-2}} \over {\cos^2(\delta t/\tau_{\rm loss})[
T_{-}(E/E_{\rm p})+\tan(\delta t/\tau_{\rm loss})/\delta]^{p_0}}},
\end{equation}
where $\delta^2 = 1-b^2/4$, $b=a{\cal D}\tau_0 E_p^2=\tau_{\rm loss}/\tau_{\rm ac}$ and
$T_{\pm} = 1 \pm b\tan(\delta
t/\tau_{\rm loss})/(2\delta)$.  Note that $b=0$ correspond to the case of no
acceleration described above. This solution is valid for $b^2<4$.  For $b^2>4$
we are dealing with an imaginary value
for $\delta$ so that tangents and cosines become hyperbolic functions with
$\delta^2 =b^2/4 - 1$. For $\delta=0$ or $b=2$ this expression reduces to
\begin{equation}\label{prompt3}
N(E,t) = \exp \{-t/T_{\rm esc}\}Q_0{{[1 - (E/E_{\rm p} -
1)t/\tau_{\rm loss}]^{p_0-2}} \over {[E/E_{\rm p}-(E/E_{\rm p} - 1)t/\tau_{\rm loss}]^{p_0}}}.
\end{equation}

The right panel of Fig.~\ref{spectra1} shows the evolution of an initial power
law spectrum subjected to weak acceleration ($b=2$, solid lines) and a fairly
strong rate of acceleration  ($b=60$, dashed line). As expected with
acceleration, one can push the electron spectra to higher levels and extend it
to higher energies. At low rates of acceleration the spectrum  evolves toward
the generic case of a flat low energy part with  a fairly steep  cutoff above
$E_{\rm p}$. At higher rates, and for some periods of time comparable to
$\tau_{\rm ac}$, the cut off energy $E_{\rm cr}$  will be greater than $E_{\rm
p}$ and there will be a power law portion below it.\footnote{At even later times
than shown here on gets a large pile up at the cut off energy (see P01). This
latter feature is of course artificial because we have neglected the diffusion
term which will smooth out such features (see \protect\citealt{Brunetti2007}).}
As evident from this figure there are periods of time when in the relevant
energy range (thick solid lines) the spectra resemble what is needed for
describing the radio and hard X-ray observations from Coma described in Fig.~6
of \citealt{Petrosian2008} - Chapter 10, this volume.

{\sl In summary, it appears that a steady state model has difficulties and that
the most likely scenario is episodic injection of relativistic particles and/or
turbulence and shocks which will re-accelerate the existing or injected
relativistic electrons into a spectral shape consistent with observations.
However these spectra are short lived, lasting for periods of less than a
billion years.}

\section{Summary and conclusion}

We have given a brief overview of particle acceleration in astrophysical plasmas
in general, and acceleration of electrons in the ICM in particular. We have
pointed out the crucial role plasma waves and turbulence play in all
acceleration mechanisms and outlined the equations that describe the generation,
cascade and damping of these waves and the coupling of these processes to the
particle kinetics and energising of the plasma and acceleration in both
relativistic and nonrelativistic regimes.

We have applied these ideas to the ICM of clusters of galaxies with the aim of
production of electron spectra which can explain the claimed hard X-ray emission
either as  non-thermal bremsstrahlung emission by nonrelativistic electrons or
as inverse Compton emission via scattering of CMB photons by a population of
relativistic electrons. It is shown that the first possibility which can come
about by accelerating background  electrons into a non-thermal tail is not a
viable mechanism, as was pointed earlier in P01. The primary reason for this
difficulty is due to the short Coulomb collision and loss timescales. Quite
generally, it can be stated that at low rates of acceleration one obtains a
hotter plasma and an insignificant non-thermal tail. Discernible tails can be
obtained at higher rates of acceleration but only for short periods of time. For
periods on the order of a billion year such rates will also cause excessive
heating and will lead to runaway conditions where most of the electrons are
accelerated to relativistic energies, at which they are no longer bound to the
cluster, unless there exists a strong scattering agent.

This leads us to the model where hard X-rays are produced by the inverse Compton
process and relativistic electrons. Moreover, even if the hard X-ray radiation
turns out to be not present, or one finds a way to circumvent the above
difficulties, we still require the presence of relativistic electrons to explain
the radio emission. These electrons must be injected into the ICM by some other
means. They can come from galaxies, specially when they are undergoing an active
nuclear (or AGN) phase. Or they may be due to interactions of cosmic ray protons
with thermal protons and the resultant pion decays. We have shown that just
injection may not be sufficient, because for reasonable injected spectra the
transport effects in the ICM modify the spectrum such that the effective
radiating spectrum is inconsistent with what is required. Thus, a
re-acceleration in the ICM is necessary and turbulence and merger shocks may be
the agent of this acceleration. In this case, it also appears that a steady
state scenario, like the hadronic mechanism described above, will in general
give a flatter than required spectrum unless the electrons escape the ICM
unhindered. This requirement is not reasonable because the expected tangled
magnetic field will increase this time. But, more importantly, the presence of
turbulence necessary for re-acceleration will result in a short mean free path
and a much longer escape time.

A more attractive scenario is if the injection of electrons and/or the
production of turbulence is episodic. For example for a short lived electron
injected phase (from say an AGN) but a longer period of presence of turbulence
one can determine the spectral evolution of the electrons subject to
acceleration and losses. We have shown that for some periods of time lasting
several times the acceleration timescales one can obtain electron spectra
consistent with what is required by observations. The same will be true for a
hadronic source if there is a short period of production of turbulence. In
either case we are dealing with periods on the order of several  hundreds of
million years to a billion years, which is comparable with timescales expected
from merging of Mpc size clusters with velocities of several thousands of
km\,s$^{-1}$ which are theoretically reasonable and agree with observations (see
e.g. \citealt{Bradac2006}).

\begin{acknowledgements} 
The authors thank ISSI (Bern) for support of the team ``Non-virialized X-ray
components in clusters of galaxies''. A.M.~B. acknowledges the RBRF grant
06-02-16844 and a support from RAS Programs.
\end{acknowledgements}

\appendix

\section{Particle kinetic equations}
\label{detail}

In this section we describe some of the mathematical details required for
investigation of the acceleration and transport of all charged particles
stochastically and by shocks, and the steps and conditions that lead to the
specific kinetic equations (Eq.~\ref{KEQ}) used in this and the previous
chapters.

\subsection{Stochastic acceleration by turbulence}

In strong magnetic fields, the gyro-radii of particles are much
smaller than the scale of the spatial variation of the field, so
that the gyro-phase averaged distribution of the particles depends
only on four variables: time, spatial coordinate $z$ along the
field lines, the momentum $p$, and the pitch angle $\cos\mu$.
In this case, the evolution of the particle distribution, $f(t, z,
p, \mu)$, can be described by the Fokker-Planck equation as they
undergo stochastic acceleration by interaction with plasma
turbulence (diffusion coefficients $D_{pp},\,D_{\mu\mu}$ and
$D_{p\mu}$), direct acceleration (with rate $\dot{p}_G$), and
suffer losses (with rate $\dot p_L$) due to other interactions
with the plasma particles and fields:
\begin{eqnarray}
{\partial f\over \partial t}+v\mu{\partial f\over \partial z} &=&
{1\over p^2}{\partial\over\partial p}p^2
\left[D_{pp}{\partial
f\over\partial p} + D_{p\mu}{\partial f\over\partial \mu}\right] \nonumber \\
&+&
{\partial \over\partial \mu} \left[D_{\mu\mu}{\partial
f\over\partial \mu} + D_{\mu p}{\partial f\over\partial p}\right] \nonumber \\
&-&
{1\over p^2}{\partial\over\partial p}[p^2(\dot{p}_L-\dot{p}_G)f] \nonumber \\
&+&
\dot{J}. 
\label{FPeq}
\end{eqnarray}

Here $\beta {\rm c}$ is the velocity of the particles and ${\dot{J}(t,z,p,\mu)}$
is a source term, which could be the background plasma or some injected spectrum
of particles. The kinetic coefficients in the Fokker-Planck equation can be
expressed through correlation functions of stochastic electromagnetic fields
(see e.g. \citealt{Melrose1980,Berezinskii1990,Schlickeiser2002}). The effect of
the mean magnetic field convergence or divergence can be accounted for by
adding 
\begin{equation}
{{\rm c}\beta {\rm d}\, \ln B\over {\rm d}s}
{\partial\over\partial\mu}\left({(1-\mu^2)\over 2} f\right) \nonumber
\end{equation} to the right hand side.

{\sl Pitch-angle isotropy:} At high energies and in weakly magnetised plasmas
with  Alfv\'en velocity $\beta_{\rm A}\equiv v_{\rm A}/{\rm c}\ll 1$ the ratio
of the energy and pitch angle diffusion rates $D_{pp}/p^2D_{\mu\mu}\approx
(\beta_{\rm A}/\beta)^2\ll 1$, and one can use the {\sl isotropic approximation}
which leads to the {\sl diffusion-convection equation} (see e.g.
\citealt{Dung1994,Kirk1988}):
\begin{equation} 
F(z,t,p) \equiv {1\over
2}\int\limits_{-1}^1{\rm d}\mu f(t,z,p,\mu), \ \ \ \ 
{\dot Q}(t,z,p) \equiv
{1\over 2}\int\limits_{-1}^1{\rm d}\mu {\dot J}(\mu, z, t, p), 
\end{equation}
\begin{eqnarray}
{\partial F\over\partial t}-{\partial\over \partial
z}\kappa_1{\partial F\over \partial z} &=& (pv){\partial
\kappa_2\over\partial z}{\partial F\over \partial p} \nonumber \\
&-&
{1\over
p^2}{\partial \over \partial p}(p^3v\kappa_2){\partial F\over
\partial z} \nonumber \\
&+& 
{1\over p^2}{\partial\over \partial
p}\left(p^4\kappa_3{\partial F\over\partial
p}-p^2\dot{p}_{\rm L}F\right)  \nonumber \\
&+& {\dot Q}(z,t,p)\,, 
\label{dceq}
\end{eqnarray}
\begin{eqnarray}
\kappa_1&=& {v^2\over 8}\int\limits_{-1}^1{\rm d}\mu{(1-\mu^2)^2\over
D_{\mu\mu}}\,,\nonumber \\
\kappa_2&=& {1\over 4} \int\limits_{-1}^1{\rm d}\mu(1-\mu^2){D_{\mu p}\over p
D_{\mu\mu}}\,,\label{kappa2}\nonumber \\
\kappa_3 &=& {1\over 2}\int\limits_{-1}^1{\rm d}\mu(D_{pp}-D^2_{\mu
p}/D_{\mu\mu})p^2\,.
\label{kappa3}\nonumber
\end{eqnarray}
At low energies, as shown by Pryadko \& Petrosian (1997), specially for strongly
magnetised plasmas ($\alpha\ll 1, \beta_{\rm A} >1$), $D_{pp}/p^2\gg
D_{\mu\mu}$, and then stochastic acceleration is more efficient than
acceleration by shocks ($D_{pp}/p^2\gg \dot{p}_G$). In this case the pitch angle
dependence may not be ignored.
\begin{eqnarray}
{\partial f^\mu\over \partial t}+v\mu{\partial f^\mu\over \partial
z}& =& {1\over p^2}{\partial\over \partial
p}p^2D^\mu_{pp}{\partial f^\mu\over\partial p} -{1\over
p^2}{\partial\over\partial
p}[p^2\dot{p}_{\rm L}f^\mu]+\dot{J}^\mu\,, \label{dceq1}
\end{eqnarray}
However, \citet{Petrosian2004} find that these dependences are
in general weak and one can average over the pitch angles.

\subsection{Acceleration in large scale turbulence and shocks}

In an astrophysical context it often happens that the energy is released at
scales much larger than the mean free path of energetic particles. If the
produced large scale MHD turbulence is supersonic and superalfv\'enic then MHD
shocks are present in the system. The particle distribution within such a system
is highly intermittent. Statistical description of intermittent systems differs
from the description of homogeneous systems. There are strong fluctuations of
particle distribution in shock vicinities. A set of kinetic equations for the
intermittent system was constructed by \citet{Bykov1993}, where the smooth
averaged distribution obeys an integro-differential equation (due to strong
shocks), and the particle distribution in the vicinity of a shock can be
calculated once the averaged function was found.

The pitch-angle averaged distribution function $N({\mbox{\boldmath $r$}},p,t)$
of non-thermal particles (with energies below some hundreds of GeV range in the
cluster case) averaged over an ensemble of turbulent motions and shocks
satisfies the kinetic equation
\begin{equation}
      \frac{\partial f}{\partial t} -
       \frac{\partial}{\partial r_{\alpha}} \: \chi_{\alpha \beta} \:
       \frac{\partial f}{\partial r_{\beta}}  =
       G  \hat{L} f +
      \frac{1}{p^2} \: \frac{\partial}{\partial p} \: p^4 D \:
      \frac{\partial f}{\partial p} + A {\hat{L}}^2 N +
      2B \hat{L} \hat{P} f + {\dot J}(p),\label{eq:ke1}
\end{equation}

The source term ${\dot J}(t,r,p)$ is determined by injection of particles. The
integro-differential operators $\hat{L}$ and $\hat{P}$ are given by

\begin{equation}
      \hat{L}= \frac{1}{3p^2} \: \frac{\partial}{\partial p} \:
      p^{3-\gamma} \: \int_{0}^{p} {\rm d}p' \: {p'}^\gamma
      \frac{\partial}{\partial p'} \;;~~~~~
      \hat{P}= \frac{p }{3} \: \frac{\partial}{\partial p} \, .
\end{equation}

The averaged kinetic coefficients $A$, $B$, $D$, $G$, and
$\chi_{\alpha \beta} = \chi \, \delta_{\alpha \beta}$ are
expressed in terms of the spectral functions that describe
correlations between large scale turbulent motions and shocks, the
particle spectra index $\gamma$ depends on the shock ensemble
properties (see \citealt{Bykov1993}). The kinetic coefficients
satisfy the following renormalisation equations:

\begin{equation}
  \chi = \kappa_1(p) + {1 \over 3} \int { {\rm d}^3 {\bf k} \, {\rm d}\omega \over (2\pi)^4 }
  \left[ {2T+S \over {\rm i}\omega+k^2\chi }
        -{2k^2\chi S \over \left( {\rm i}\omega + k^2\chi\right)^2 } \right]
        \,,
        \label{eq:Chi}
\end{equation}

\begin{equation}
  D={\chi \over 9} \int { {\rm d}^3 {\bf k} \,  {\rm d}\omega \over (2\pi)^4 }\;
   {k^4 S(k,\omega) \over \omega^2 + k^4 \chi^2 } \, , \label{eq:D}
\end{equation}

\begin{equation}
 A = \chi \int { {\rm d}^3 {\bf k} \, {\rm d}\omega \over (2\pi)^4 } \;
   {k^4 \tilde{\phi}(k,\omega) \over \omega^2 + k^4 \chi^2 }\, , \label{eq:A}
\end{equation}

\begin{equation}
  B = \chi \int { {\rm d}^3 {\bf k}\, {\rm d}\omega \over (2\pi)^4 }\;
  {k^4 \tilde{\mu}(k,\omega) \over \omega^2 + k^4 \chi^2 }\, .  \label{eq:B}
\end{equation}
Here $G = ( 1/\tau_{\rm sh}+ B)$.  $T(k,\omega)$ and $S(k,\omega)$ are the
transverse and longitudinal parts of the Fourier components of the turbulent
velocity correlation tensor. Correlations between velocity jumps on shock fronts
are described by $\tilde{\phi}(k,\omega)$, while $\tilde{\mu}(k,\omega)$
represents shock-rarefaction correlations. The introduction of these spectral
functions is dictated by the intermittent character of a system with shocks.

The test particle calculations showed that the low energy branch of the particle
distribution would contain a substantial fraction of the free energy of the
system after a few acceleration times. Thus, to calculate the efficiency of the
shock turbulence power conversion to the non-thermal particle component, as well
as the particle spectra, we have to account for the backreaction of the
accelerated particles on the shock turbulence. To do that, \citet{Bykov2001}
supplied the kinetic equations Eqs.~\ref{eq:ke1})$-$(\ref{eq:B} with the energy
conservation equation for the total system including the shock turbulence and
the non-thermal particles, resulting in temporal evolution of particle spectra.

{}

\end{document}